\documentclass[12pt,preprint]{aastex}
\usepackage{emulateapj5}

\slugcomment{Accepted For Publication In ApJ}

\shorttitle{Ly$\alpha$ Emitters at $z=4.86$}
\shortauthors{Ouchi et al.}

\begin{document}


\title{
Subaru Deep Survey II.
Luminosity Functions and Clustering Properties of
Ly$\alpha$ Emitters at $z=4.86$
in the Subaru Deep Field \altaffilmark{1} }


\author{Masami Ouchi        \altaffilmark{2},
	Kazuhiro Shimasaku  \altaffilmark{2,3},
	Hisanori Furusawa   \altaffilmark{2},
	Masayuki Miyazaki   \altaffilmark{2},\\
	Mamoru Doi          \altaffilmark{3,4},
	Masaru Hamabe       \altaffilmark{5},
	Tomoki Hayashino    \altaffilmark{6},
	Masahiko Kimura     \altaffilmark{7},\\
	Keiichi Kodaira	    \altaffilmark{8},
	Yutaka Komiyama     \altaffilmark{9},
	Yuichi Matsuda      \altaffilmark{6},
	Satoshi Miyazaki    \altaffilmark{10},\\
	Fumiaki Nakata      \altaffilmark{2},
	Sadanori Okamura    \altaffilmark{2,3},
	Maki Sekiguchi      \altaffilmark{7},
	Yasuhiro Shioya     \altaffilmark{11},\\
	Hajime Tamura       \altaffilmark{6},
	Yoshiaki Taniguchi  \altaffilmark{11},
	Masafumi Yagi       \altaffilmark{10}, and
	Naoki Yasuda        \altaffilmark{10}
	}

\email{ouchi@astron.s.u-tokyo.ac.jp}


\altaffiltext{1}{Based on data collected at
	Subaru Telescope, which is operated by
	the National Astronomical Observatory of Japan.}
\altaffiltext{2}{Department of Astronomy, School of Science,
         University of Tokyo, Tokyo 113-0033, Japan}
\altaffiltext{3}{Research center for the Early Universe, School of Science,
         University of Tokyo, Tokyo 113-0033, Japan}
\altaffiltext{4}{Institute of Astronomy, School of Science,
	University of Tokyo, Mitaka, Tokyo 181-0015, Japan}
\altaffiltext{5}{Department of Mathematical and Physical Sciences,
	Faculty of Science, Japan Women's University, Tokyo 112-8681, Japan}
\altaffiltext{6}{Research Center for Neutrino Science,
	Graduate School of Science,
	Tohoku University, Aramaki, Aoba, Sendai 980-8578, Japan}
\altaffiltext{7}{Institute for Cosmic Ray Research,
	University of Tokyo, Kashiwa, Chiba 277-8582, Japan}
\altaffiltext{8}{The Graduate University for Advanced Studies,
	Shonan village, Hayama, Kanagawa 240-0193, Japan}
\altaffiltext{9}{Subaru Telescope, National Astronomical Observatory,
	650 N.A'ohoku Place, Hilo, HI 96720, USA}
\altaffiltext{10}{National Astronomical Observatory,
	Mitaka, Tokyo 181-8588, Japan}
\altaffiltext{11}{Astronomical Institute, Graduate School of Science,
         Tohoku University, Aramaki, Aoba, Sendai 980-8578, Japan}


\begin{abstract}
  We report on early results from
very deep and wide-field narrow-band
imaging on a 543 arcmin$^2$ area of the Subaru Deep Field.
We find 87 Lyman $\alpha$ emitters (LAEs) at $z=4.86\pm 0.03$
which are photometrically
selected by a combination of two broad bands ($R$ and $i'$)
and one narrow band
($NB711$; $\lambda_{c}=7126$\AA, $\Delta \lambda = 73$\AA).
We derive the luminosity functions (LFs) of the LAEs at Ly$\alpha$ luminosity
and at UV-continuum (rest-frame 1700\AA) luminosity.
The LFs show little evolution between
z=3.4 and z=4.86 either in Ly$\alpha$ or
UV-continuum emission. The amplitude of the LAE LF
tends to decline at the bright magnitudes more rapidly
than that of the LBG LF at similar redshifts.
We calculate the angular correlation function of our LAEs
up to $\sim15$ arcmin separations.
The angular correlation $\omega(\theta)$ is found to
increase with decreasing angular separations, showing 
a clear signal of clustering.
It is also found that the distribution of LAEs shows 
a large density gradient with a scale of $\gtrsim 15$ arcmin, which 
would indicate the existence of a large-scale structure of LAEs 
on $\gtrsim 20 h^{-1}$Mpc scales.
We fit the observed correlation function by
$A_\omega \theta^{-0.8}$ to find $A_\omega=29$ arcsec$^{0.8}$.
The estimated correlation length is
$r_0=3.5^{+0.3}_{-0.3}h^{-1}$Mpc in comoving units
($\Omega_m=0.3$ and $\Omega_\Lambda=0.7$), which
is slightly larger than the value for $z\sim4$ LBGs with $i'<26$.
We calculate the angular correlation function for
two sub-samples of the 87 LAEs divided by Ly$\alpha$
luminosity, UV-continuum luminosity, and Ly$\alpha$ equivalent
width ($EW$).
The Ly$\alpha$-bright sub-sample shows
a larger correlation amplitude than the Ly$\alpha$-faint sub-sample,
while no significant difference is found for the sub-samples divided by
UV-continuum luminosity or $EW$.
This may indicate that galaxies with bright
Ly$\alpha$ emission are possibly biased against the
underlying dark matter halos more strongly than those with bright
UV continuum.
\end{abstract}





\section{INTRODUCTION}
\label{sec:introduction}
  In the typical hierarchical clustering scheme
of cold dark matter (CDM) models, sub-galactic clumps are formed
from gravitational collapse of small initial density
perturbations, and those clumps progressively merge and
collapse to become galaxies of larger masses.
Merging processes increase (decrease) the number density
of bright (faint) galaxies, and enhance the clustering
amplitudes of galaxies. The luminosity functions (LFs) and
clustering properties of high-$z$ ($z>2$) galaxies
are, thus, essential for understanding the galaxy formation.
Recent observational efforts at optical wavelengths
have revealed two major classes of high-$z$ galaxies which
are suitable for studying LFs and clustering
properties of distant galaxies.

One class of high-$z$ galaxies
is Lyman break galaxies (LBGs; \citealp{steidel1996a})
which are selected by their UV-continuum features
redshifted to optical wavelengths
\footnote{
One can select high-$z$ galaxies by
photometric redshift techniques (e.g.,
\citealp{connolly1995,sawicki1997,fernandez-soto1999,
furusawa2000,fontana2000}).
However, since photometric redshift techniques also
use essentially the Lyman break feature for selecting them,
we refer to, for simplicity,
those selected by photometric redshifts as LBGs
throughout this paper.
}.
Since LBGs are identified by their strong
continuum break, they
are generally limited to UV-continuum bright galaxies.
Such bright high-$z$ galaxies can be
well investigated. 
The LF of LBGs at $z\sim3$ \citep{steidel1999}
has a larger $M^*$
than that of the local UV-selected galaxies 
\citep{sullivan2000},
but its normalization, $\phi^*$, is 
comparable within a factor of two
for the LBGs and for the local galaxies
(see also \citealp{lowenthal1997}).
The apparent size of LBGs at $z\sim3$ is $0''.2$-$0''.3$
($1.0-1.6 h^{-1}$kpc
for $\Omega_m=0.3$ and $\Omega_\Lambda=0.7$)
in a half light radius \citep{steidel1996b}.
LBGs at $z\sim3-4$ show stronger clustering than
the underlying dark matter \citep{giavalisco1998,ouchi2001a}.
A hint of mergers has been
found for LBGs at $z\sim4$ \citep{ouchi2001a}.

The other class of high-z galaxies
is Lyman $\alpha$ Emitters (LAEs)
which are identified by their
redshifted Ly$\alpha$ emission
with narrow-band imaging
\citep{cowie1998,hu1998}.
LAEs are thus
high-$z$ galaxies which have
a relatively strong Ly$\alpha$ emission
whose rest-frame equivalent width,
$EW_{\rm rest}$, is $\gtrsim 20$\AA.
Most of the detected LAEs tend to have
fainter continua than LBGs (e.g.,
\citealp{cowie1998,hu1998,steidel2000,fynbo2001}).
The estimated number density of LAEs
down to $f_{\rm Ly\alpha} = 2-5 \times
10^{-17} {\rm erg\ s^{-1} cm^{-2}}$ is
$\sim 10,000$deg$^{-2}$ per unit $z$
\citep{hu1998,rhoads2000}.
The size of LAEs at $z=2.4$ is as small as
$\lesssim 0''.1$
($\lesssim 0.6 h^{-1}$kpc
for $\Omega_m=0.3$ and $\Omega_\Lambda=0.7$)
in a half light radius \citep{pascarelle1996}.
An indication for a large scale       
structure formed by LAEs at $z=2.4$           
has been reported by \citet{pascarelle1998}
from a combination of their three blank field surveys, 
although other blind searches have not found such
clumpy distributions \citep{rhoads2000}.
Searches of LAEs on targetted fields have found strong
clustering of LAEs around quarars 
\citep{campos1999,moller2001,stiavelli2001},
radio galaxies \citep{venemans2002},
high redshift clusters \citep{palunas2000},
and overdensities in the LBG distribution 
\citep{steidel2000}.

Since LAEs are fainter and smaller than LBGs
and some LAEs have strong clustering, 
it is possible that LAEs are dominated by
sub-galactic clumps which subsequently
form massive galaxies.
Indeed, \citet{pascarelle1996} have proposed
that LAEs are sub-galactic clumps.
If high-$z$ LAEs
are dominated by sub-galactic clumps,
the LF of LAEs is expected to show
a steep decline toward the bright end.
The clustering amplitude of LAEs should be
larger than that of LBGs on small
scales ($\lesssim 20$kpc or $\lesssim 5''$)
where the correlation amplitude is sensitive
to the fraction of merging galaxies.

In this paper, we report on luminosity functions
and clustering properties of LAEs at $z=4.86$, based on
our deep ($i'\sim27$) and wide-field ($540$arcmin$^2$)
broad- and narrow-band imaging of the Subaru Deep Field.
The unprecedentedly deep and wide-field
imaging data enable us to investigate
the LF of LAEs down to $M_{\rm UV}\sim-18.5$
and the clustering properties up to $\sim20$ Mpc scales
in comoving units.
Throughout this paper, magnitudes are in the AB system, and
all calculations assume a $\Lambda$-dominated spatially flat
cosmology, $H_0=100 h {\rm km s^{-1} Mpc^{-1}}$,
$\Omega_m=0.3$, and $\Omega_\Lambda=0.7$.

\section{OBSERVATIONS AND DATA REDUCTION}
\label{sec:observations_and_data_reduction}
As part of the Subaru Deep Field (SDF) project,
deep and wide-field optical imaging
of the SDF
($13^{\rm h} 24^{\rm m} 21^{\rm s}.38$, $+27^\circ 29' 23''.0$[J2000])
was carried out during the commissioning phase of
Subaru Prime Focus Camera
(Suprime-Cam; \citealp{miyazaki1998})
in March to June, 2001
(Shimasaku et al. in preparation).
We obtained $B$-, $V$-, $R$-, $i'$-, $z'$- and a narrow-band ($NB711$:
centered at 7126\AA) images of
about a 600 arcmin$^2$ area of the SDF
\footnote{
The central 4 arcmin$^2$ region of SDF
has ultra-deep $J$ and $K'$ data \citep{maihara2001}
}.
Figure \ref{fig:sed_response}a shows the transmittance of
$R$,$i'$, and $NB711$.
The peak transmittance, central wavelength, and band width (FWHM)
of $NB711$ are measured to be
$0.83\pm 0.01$, $7126\pm 4$\AA, and $73.0\pm 0.6$\AA,
respectively. The errors indicate
the rms measured at
13 positions evenly spaced over the surface
of the filter.
The data were reduced in the same manner as explained in
\citet{ouchi2001a}. The final images cover a contiguous
543 arcmin$^2$ area with a PSF FWHM of $0.''90$.
The net exposure times of the final images are
210, 150, 90, 138, 81, and 162 minutes for $B$, $V$, $R$, $i'$, $z'$,
and $NB711$, respectively. The limiting AB magnitudes are
$B=27.8$, $V=27.3$, $R=27.1$, $i'=26.9$, $z'=26.1$,
and $NB711=26.0$ for a 3$\sigma$ detection on a
$1.''8$ diameter aperture.
Source detection and photometry are performed using
SExtractor version 2.1.6 \citep{bertin1996}.
The $NB711$-band frame is chosen to detect objects,
and we limit the object catalog to
$NB711<26.0$, in order to provide
a reasonable level of photometric completeness.

\section{PHOTOMETRIC SAMPLE OF $z=4.86$ LY$\alpha$ EMITTERS}
\label{sec:selection_sample}

Our catalog contains 30,297 objects
with $NB711<26.0$ in total. Figure \ref{fig:cm} shows
the color magnitude diagram of the detected objects. Since
the $NB711$ band measures fluxes between the
$R$ and $i'$ bands (see Figure \ref{fig:sed_response}a),
we define the off-band continuum flux of objects
as $Ri\equiv(R+i')/2$.
We determine the selection criteria
which isolate LAEs at $z=4.86\pm 0.03$ from
foreground objects,
on the basis of expectations from SEDs of model LAEs,
foreground galaxies, and Galactic stars (left panel of
Figure \ref{fig:cc}; see the figure caption for the details).
We regard an object as a LAE when it satisfies the
following selection criteria simultaneously,

\begin{eqnarray}
Ri-NB711>0.8,\\
R-i'>0.5,\\
i'-NB711>0.
\label{eq:laeselection}
\end{eqnarray}

  The first criterion corresponds to an equivalent
width ($EW$) to be $>82$\AA\ in the observed frame
( $>14$\AA\ in the rest frame at $z=4.86$)
\footnote{
For simplicity, we calculate EW from $Ri-NB711$ alone,
without correcting for $R-i$ color.
}.
The second criterion rejects low-$z$ objects
with strong emission lines of
H$\alpha$, [OIII], and [OII] etc. Since the
Ly$\alpha$ continua of high-$z$ objects are
absorbed by neutral hydrogen of the
inter-galactic medium (IGM), $R$-band fluxes are
expected to be
weaker than $i'$-band fluxes for $z=4.86$ objects.
Figure \ref{fig:cc} demonstrates that
this criterion efficiently rejects low-$z$ emitters
but selects $z=4.86$ galaxies whose IGM absorptions
are stronger than 0.5 $ \tau_{\rm eff}$, where
$ \tau_{\rm eff}$ is the median opacity of the IGM given by
\citet{madau1995}.
The third criterion is placed to reject
break objects whose spectra have a spectral trough
on the bluer side of the $NB711$-band. This
trough is misidentified as an emission feature
by $Ri-NB711$ color.
These three criteria collectively select
LAEs at $z=4.86$ with a low
contamination from low-$z$ interlopers.

There are a total of 87 objects that meet all the criteria 
simultaneously down to $NB711=26.0$ and $i'=27.8$, where 
$i'=27.8$ is the $1.3\sigma$ limiting magnitude of the i' band.
The color criterion, $R-i'>0.5$, can be securely
imposed on objects detected over a $3\sigma$ level in $i'$ ($i'=26.9$)
by using their $R$ magnitudes
or a $2\sigma$ upper limit ($R=27.6$).
Among the 87 selected objects, 62 objects brighter than 
$i'=26.9$ have a secure $R-i'>0.5$ limit.
Remaining 25 ($=87-62$) objects with $i'\geq 26.9$ have large errors 
in $R-i'$ color, and thus the contamination by foreground objects 
will be higher than for the bright objects.
However, in order to enlarge our sample, we do not remove 
these faint objects from our LAE sample, and insted estimate 
the contamination and completeness for the whole sample 
as a function of apparent magnitude on the basis of Monte 
Carlo simulations which take account of photometric errors.
Our photometric sample of $z=4.86$ LAEs comprizes these 87 objects.
These LAEs are shown by large red points in Figure \ref{fig:cm}.

In Figure \ref{fig:cm}, some objects seem to have absorption features 
at significant levels. We do not know the detail spectral
features of those objects, but for simplicity
we call those objects with absorption features "absorbers". 
We define as absorbers those having  
$Ri-NB711 <-0.6$ and $Ri-NB711<-0.4NB711+9.6$
(yellow line in Figure \ref{fig:cm}), and find 37 absorbers.
We investigate the SEDs of the 37 absorbers from $BVRi'z'$ multi-color
data. Fifteen objects show a good agreement with stellar spectra,
and eight objects are consistent with late-type galaxies.
However, the nature of the remaining fourteen objects is uncertain.
Their spectra marginally
agree with spectra of low-$z$ galaxies or Galactic objects.
They may be variable objects and/or moving
objects (such as Kuiper-belt objects).
Future spectroscopic follow-up observations
will reveal the nature of those objects
\footnote{Our on-going spectroscopic
obaservations have revealed that 
one absorber is a $z=0.58$ galaxy with a very blue continuum
and strong [OIII] emission lines in the $i'$-band.}

We have estimated photometric
contamination and completeness of the LAE sample.
Since the sample
construction is based on $NB711$, $R$, and $i'$
images, we estimate the completeness and contamination
by Monte Carlo simulations using real $R$-,$i'$-,
and $NB711$-band images.
To estimate the contamination,
we generate $30,297$ objects that mimic all the detected objects
with $NB711<26$, and distribute them randomly on our original images
after adding Poisson noises according to their original brightness.
Then, we detect these simulated objects and measure their brightness.
The contamination of our sample is defined by the ratio of the
objects that did not pass the criteria in the original
data but satisfy the criteria in the simulated data,
to the number of all the original objects which passed
the criteria. We iterate this process 10 times and
estimate the ratio of contamination as a function of 
$i'$ magnitude. We find the sample contamination 
to be 40\%.
We calculate the completeness of LAEs as a function of 
continuum ($i'$) magnitude, not as a function of NB711 
magnitude, since we want to derive the rest-frame 
UV-continuum luminosity function of LAEs (see section \ref{sec:luminosity_function}). 
We need to take two kinds of incompleteness into account; 
one is the incompleteness of $i'$-band detection, 
and the other is the incompleteness due to our sample 
being $NB711$-limited.
The latter incompleteness is equivalent to missing 
small-$EW$ emitters of faint $i'$ magnitudes
\footnote{
In terms of $NB711$-dependent incompleteness, 
our $NB711$-limited sample of LAEs may miss large-$EW$ emitters 
of faint $NB711$ magnitudes, since the limiting magnitude 
of $NB711$ and $i'$ for our sample selection is $NB711=26.0$ and 
$i'=27.8$.
This implies that objects with $i'-NB711\gtrsim 2$ will not be 
included in the sample at the faintest $NB711$ magnitudes
($NB711\sim 26.0$). Since we cannot determine a subsample 
which is complete in terms of emission luminosity, 
we do not calculate completeness as a function of 
emission luminosity in our analysis.
}.
To estimate the completeness (the sum of the above two 
sources), we make a subsample composed of 45 LAEs with 
$i'<26.5$, whose continuum flux and 
$EW$ (or $Ri-NB711$ color) are fairly accurately measured. 
This $i'$-limited subsample is essentially complete 
in terms of $EW$ (or $Ri-NB711$ color).
Assuming that the $R-i'$ vs. $Ri-NB711$ color distribution 
found for the objects in this subsample is universal, i.e.,
irrespective of $i'$ magnitude, we make a mock catalog of 
360 ($=45 \times 8$) LAEs whose $i'$-band magnitudes
are scaled 
from $i'=24.5$ to $i'=27.5$ into 0.5 magnitude bins.
We perform Monte Carlo simulations with 360 artificial LAEs 
in the same manner as for the contamination estimation.
We define here the completeness of LAE sample as the 
number of the simulated objects which again pass 
the selection criteria divided by all the original objects
which passed the criteria.
We repeat this simulation 100 times, and average 
the measured completenesses. 
The completeness weighted by 
the number of the detected objects are found 
to be 40\%
\footnote{
These values of contamination and completeness 
include photometric errors only,
and do not include any systematic errors.
The systematic errors may be caused from
(1)low-$z$ interlopers satisfying
the selection criteria and
(2)the photometric errors of input objects
themselves by which the observed
color distributions are wider
than the real distributions
in the $Ri-NB711$ vs. $R-i'$ plane.
More reliable contamination and completeness estimations
can, however, be obtained only with spectroscopic follow-up
observations.}.
When deriving the UV luminosity function in 
section \ref{sec:luminosity_function}, 
we correct the data for the contamination and completeness 
derived in this way.
The completeness correction derived here is valid 
if the color distribution of LAEs does not change largely 
with apparent $i'$ magnitude. 

In order to examine how well eqs. (1)-(3) can select
$z=4.86$ galaxies, we
examine what fraction of
LAEs satisfy
continuum break selection criteria
for $z\sim5$ LBGs.
On the basis of expectations from
GISSEL96 \citep{bruzual1993}
population synthesis models,
we define the criteria for $z\sim5$ LBGs as
$V-i'>1.8$, $i'-z'<0.7$, and $V-i'>2.0 \times (i'-z')+1.4$,
which isolate $z\sim5$ galaxies from low-$z$ galaxies
and Galactic stars (
Ouchi et al. in preparation;
see also \citealp{ouchi2001b}).
Since most of our LAEs have faint continuum
magnitudes (i.e., $i'\sim27$),
the number of LAEs to which the LBG selection
criteria can be reliably applied is very limited.
This is because the LBG selection criteria
require very deep $V$ magnitudes
relative to $i'$ magnitudes.
The limiting magnitudes of our $V$ data
are $\simeq27.8$ in a $2\sigma$ level,
and thus the LBG criteria can be
safely applied only to galaxies brighter
than $i'\sim26.0$.
We have nine LAEs brighter than $i'=26$, and find that
five out of the nine to pass the
LBG criteria, implying that the contamination of
our LAE sample by foreground objects is $\sim 40\%$.
This fraction of contaminants is consistent with that
derived by the Monte Carlo simulations above.
On the other hand, although we cannot make the same test to
the remaining faint LAEs,
we find that no object fainter than $i'=26$ violates
clearly (i.e., above 2$\sigma$ levels) the criterion, $V-i'>1.8$.
 From these tests, we conclude that statistical properties of LAEs
can be reliably derived from our sample.

\section{Luminosity Functions}
\label{sec:luminosity_function}
We derive the luminosity function (LF) of LAEs
in Ly$\alpha$ and UV (rest-frame $1700$\AA) luminosities,
assuming that the surveyed volume is
approximated by the box with the cross section of
surveyed area and the depth
corresponding to the $NB711$'s
band width (i.e., FWHM).
Figures \ref{fig:lumifun}a and \ref{fig:lumifun}b
plot the Ly$\alpha$ LF and UV LF,
together with those of $z=3.4$
LAEs by \citet{cowie1998}.
\citet{cowie1998} selected all objects
with observed equivalent width of
$EW_{\rm obs}>77$\AA\
($EW_{\rm rest}>17.5$\AA\ in the rest frame).
Their criterion
corresponds to $EW_{\rm obs}>86$ ($Ri-NB711>0.83$)
for $z=4.86$ LAEs, if
the cosmological effects
and the difference in absorption by IGM
neutral hydrogen \citep{madau1995}
at $z=3.4$ and $z=4.86$ are taken into account.
In order to compare the
LF at $z=3.4$ with that at $z=4.86$, we make another
sample for $z=4.86$ LAEs which is selected
by the same criterion as of \citet{cowie1998}
($Ri-NB711>0.83$) from our
all detected objects. We find that 234 objects
satisfy this criterion.
We show the
LFs of these 234 objects by inverse open triangles
in Figure \ref{fig:lumifun}.
These LFs are not corrected for the sample
contamination or completeness, and they
are probably contaminated by low-$z$ emission-line galaxies,
similar to \citeauthor{cowie1998}'s \citeyearpar{cowie1998}
LF for $z=3.4$ LAEs.
Both LFs decline at the faint end due to incompleteness,
which is not corrected.

In Figure \ref{fig:lumifun}a, we calculate
the Ly$\alpha$ luminosity from
$Ri-NB711$ color and $Ri$ magnitude, correcting for
the absorption by neutral hydrogen in IGM.
A similar correction has been made
to the $z=3.4$ LAE.
Figure \ref{fig:lumifun}a shows that the
Ly$\alpha$ LF at $z=4.86$ is very similar to that at $z=3.4$.
No clear evolution is found between $z=3.4$ and $z=4.86$
within a factor 2-3.
Note that both Ly$\alpha$ LFs are likely to be 
highly contaminated by foreground low-$z$ emitters.
The pure equivalent width selection applied to our data
($Ri-NB711 \gtrsim 0.8$) finds 234 objects, while
our color selection finds only 87 objects. 
This implys that as large as 60\% of the galaxies contributing 
to the LF of $z=4.86$ LAEs in Figure \ref{fig:lumifun}a may be contaminants.
A similarly large fraction of contaminants may be expected 
for the LF of $z=3.4$ LAEs.
Thus, an actual uncertainty in our finding that no evolution 
is seen for LAEs between $z=3.4$ and 4.86 can be within a
factor of 2-3, which is significantly larger than estimated from
Figure \ref{fig:lumifun}a.

In Figure \ref{fig:lumifun}b,
the 1700 \AA\ continuum absolute magnitude
$M_{1700}$ is estimated from $i'$ magnitudes
\footnote{
Most of the LAEs have PSF-like (FWHM$=0''.9$)
shapes. If the fluxes of PSF objects
are measured with a $1''.8$ circular aperture,
they are fainter than the total magnitudes by 0.2mag.
When deriving the luminosity function in Figure 4,
we adopt $m(1.''8)-0.2$ as the total magnitude of
a LAE.
}.
Since the $i'$-band measures the flux at the
rest-frame 1300\AA\ for a $z=4.86$ galaxy,
we add the median color, $m_{1700}-i'$, of
$z_{\rm phot}=4.8\pm0.3$ HDF galaxies
($-0.04$ mag; \citealp{furusawa2000}) to all
the $i'$ magnitudes to estimate the magnitude at 1700\AA.
We calculate the best estimated LF
of our original 87 LAEs satisfying
eqs.(1)-(3) (filled squares),
correcting for the contamination and
completeness (section \ref{sec:selection_sample}).
No correction is applied to either the LF of $z=3.4$ LAEs
(circles) or that of $234$ $z=4.86$ LAEs (inverse triangles)
which are selected by
$EW_{\rm obs}>86$.
The simple criterion, $EW_{\rm obs}>86$, gives
an unrealistically high number density
of LAEs at $M_{1700}<-21$.
This is probably due to the contamination of low-$z$
interlopers because of the lack of the selection criterion
based on $R-i'$ color.
For the reader's eye guide, we draw the LFs of
LBGs at $z\sim3$ \citep{steidel1999} and $z\sim5$
(Ouchi et al. in preparation) in Figure \ref{fig:lumifun}b.
We find that the bright-end slope of the LF of
$z=4.86$ LAEs is steeper than those of $z\sim3$ and $\sim5$
LBGs.

Both $z=5$ LBGs and $z=4.86$
LAEs seem to have a similar number density at the faintest 
magnitudes.
However, at these magnitudes, the LF of $z=5$ LBGs has errors 
of a factor of 1.5-2, and the amplitude of the LF of LAEs may 
have an offset of a factor of two or more due to 
uncertainties in the estimation of contamination, 
completeness and the surveyed volume.
Hence, we cannot conclude clearly that LAEs constitute 
a majority of high-redshift galaxies at these faint magnitudes.
This is why we discuss only the relative shape of the LF 
between $z=5$ LBGs and $z=4.86$ LAEs.
We think that there exists one possibility
that alters the shape of our LAE LF. We correct the
completeness, assuming that the distribution of EW
is the same irrespective of magnitude.
If, however, faint LAEs tend to have weak EWs, 
our completeness-corrected number density will 
be overestimated. This may make the apparent shape 
of the LF steeper.

\section{Clustering Properties}
Figure \ref{fig:skydistribution} shows the sky distribution 
of the $z=4.86$ LAEs in our sample. 
We find in this figure an inhomogeneous distribution 
of the LAEs.
Prior going to investigate the clustering properties of LAEs,
we examine the spatial homogeneity of our data.
We find that the photometric zero points of our data are
accurate within 0.1 mag over the whole field of view,
since we find that
PSF-like objects make a single sharp stellar
locus in the two-color ($R-NB711$ vs. $NB711-i'$) plane.
Then we examine possible spatial differences in the
source detection efficiency
by (1) examining the number densities of $NB711$-band detected
objects in small ($10'\times 10'$) areas covering the
survey region,
(2) measuring the limiting magnitudes
in 2700 small ($40''\times40''$) areas for each of
the $R$, $i'$, and $NB711$ images, and (3) estimating the detection
completeness of LAEs from Monte-Carlo simulations
in the same manner as in \citet{ouchi2001a}
but assuming model LAE spectra (see caption of Figure
\ref{fig:cc}).
We find no inhomogeneity in (1)-(3),
which indicates that the detection efficiency is
uniform at least on $40''$ scales.

We derive the angular two-point correlation function
$\omega(\theta)$ using the estimator defined
by \citet{landy1993}. The random sample for the estimator
is composed of 100,000 sources with the same geometrical
constraints as of the data sample. 
We fit a single power law,
$\omega(\theta)=A_\omega\theta^{-\beta}$, to the data
points, but the best fit value of $\beta$ is very 
shallow ($\beta=0.1$). 
This value is much smaller than the best fit values
of the local galaxies,$\beta\sim0.8$, and
$z=3$-$4$ LBGs, $\beta\sim 0.3$-$1.2$ 
\citep{giavalisco1998,giavalisco2001,porciani2002}.
We think that the $\beta$ value cannot be determined with an 
enough accuracy from our data, and fix $\beta$ to the 'fiducial' 
value, $\beta=0.8$. 
We fit $A_\omega$ alone to the data with
a fixed slope, $\beta \equiv 0.8$, then 
the best-fit value of A$_\omega$ is 29 arcsec$^{0.8}$.
The integral constant ($IC$)
for $\beta \equiv 0.8$ is calculated to be 0.195.
The resulting angular correlation function
for the LAEs is shown in Figure \ref{fig:acorr} after
the application of the $IC$.
We find that the reduced $\chi^2$ value of this $\beta=0.8$ fitting, 
6.7, is 1.9 times larger than that for the two-parameter 
fitting, $\chi^2/N=3.5$. Although the $\chi^2$ fitting with $\beta=0.8$ is
worse than the one with $\beta=0.1$, we cannot accurately 
derive $\beta$ from our data in any case. 
Thus we decide to calculate $A_w$ at $\beta=0.8$.
We calculate the angular correlation function
under $\sim 15$ arcmin scales, since systematic errors increase 
largely over $\sim 15$ arcmin scales, which are comparable 
to a half of the surveyed field of view.
However, we find that the sky distribution of LAEs
(Figure \ref{fig:skydistribution}) has a large gradient 
in the number density from 
the north to the south. This may imply
the existence of a large-scale structure 
whose scale could be over $15$ arcmin ($\sim 20 h^{-1}$Mpc).
An excess of the angular correlation function (Figure \ref{fig:acorr}) 
over the power law at around $300''$ will be (at least partly)
due to this large-scale structure.

  In order to investigate the dependence of clustering
amplitude on physical properties of LAEs,
we calculate the angular correlation function
by dividing the 87 LAEs
into two sub-samples
based on three properties; Ly$\alpha$ luminosity, UV magnitude,
and observed equivalent width ($EW_{\rm obs}$).
For each of the three properties, we
divide the original sample at
an appropriate boundary
in order to keep a reasonable number of objects
in each of the sub-samples.
Figure \ref{fig:acorr_seg}a shows the results
when the LAE sample is divided by Ly$\alpha$ luminosity,
$\log L_{\rm Ly\alpha}>42.2$ (30 objects : filled circles)
and $\log L_{\rm Ly\alpha}\leq 42.2$ (57 objects : open circles).
Figures \ref{fig:acorr_seg}b and \ref{fig:acorr_seg}c
are the same plots, but divided by
UV magnitudes
($M_{1700}>-19.4$ [30 objects: filled circles],
$M_{1700}\leq -19.4$ [57 objects: open circles])
and by $EW_{\rm obs}$ (
$EW_{\rm obs}>270$\AA\ [30 objects: filled circles],
$EW_{\rm obs}\leq270$\AA\ [57 objects: open circles]).
The correlation amplitude
of the Ly$\alpha$-bright sample is significantly
larger than that of the Ly$\alpha$-faint sample.
On the other hand, such a significant difference is not found
in either UV-magnitude sub-samples or $EW_{\rm obs}$ sub-samples
\footnote{
We make similar analyses, varying the boundaries of
Ly$\alpha$ luminosity, UV magnitude, and $EW_{\rm obs}$, but
keeping the number ratio of objects in the two sub-samples
(e.g., Ly$\alpha$ bright and Ly$\alpha$ faint sub-samples)
being in an appropriate range,
3:7 to 7:3. We find clear dependence of clustering amplitude
on Ly$\alpha$ luminosity in a wide range of the number ratio,
but find no such significant dependence in either UV-magnitude
sub-samples or $EW_{\rm obs}$ sub-samples.
}.
Since the Ly$\alpha$ luminosity is estimated from
the narrow-band magnitude,
the Ly$\alpha$-bright sample may be biased toward
LAEs whose Ly$\alpha$
emissions are efficiently detected
around the center of the $NB711$ band,
resulting in Ly$\alpha$-bright objects are
distributed in a smaller redshift range
than Ly$\alpha$-faint objects.
A smaller redshift
distribution gives stronger angular correlation.
If this is the case, however, a similar effect should
also be seen in the $EW_{\rm obs}$-dependent correlation function.
We do not see such effect in the $EW_{\rm obs}$ sub-samples.
Different degrees of contamination of low-$z$ interlopers
in the two Ly$\alpha$ sub-samples could cause
a segregation in correlation amplitude.
In Figure \ref{fig:acorr_seg}a,
the difference in correlation amplitude
in the sub-samples is about a factor of two.
If the Ly$\alpha$ segregation
is solely due to the contamination,
the two sub-samples must have largely different
degrees of contamination by
more than a factor of two.
Since such a large magnitude-dependent
contamination is not found in the Monte Carlo
simulations (section \ref{sec:selection_sample}),
the contamination does not seem to cause
the Ly$\alpha$-luminosity segregation.
We therefore conclude that the difference in
the correlation amplitude between the Ly$\alpha$
sub-samples is probably real.

\section{DISCUSSION \& CONCLUSIONS}
\label{sec:discussion}
  The LFs of LAEs show little
evolution between $z=3.4$ and $z=4.86$ either for
Ly$\alpha$ emission or UV continuum emission (
Figures \ref{fig:lumifun}a and \ref{fig:lumifun}b).
Both Ly$\alpha$ luminosity and UV-continuum luminosity
are sensitive to
star-formation activities and extinction
by shading HII regions.
In addition, Ly$\alpha$
luminosity is dependent on
gas density and dust composition
in HII regions, since
the Ly$\alpha$ photons are produced
by a number of
the resonance scatterings in HII regions.
Therefore, little evolution of the LAE LF
between $z=3.4$ and $4.86$ may imply that
$z=3.4$ and $z=4.86$ LAEs have
similar gas densities and dust components
in their HII regions, unless the number density
of LAE is largely changed.

Figure \ref{fig:lumifun}b shows that the LF of LAEs 
at $z=4.86$ measured in UV
continuum seems to be steeper in the bright end than the LF
of LBGs at similar redshifts, indicating that galaxies
with a stronger Ly$\alpha$ emission tend to be less luminous
in UV continuum. \citet{shapley2001} have found that 
$z=3$ LBGs with old stellar populations emit a
stronger Ly$\alpha$ emission than those with young stellar populations.
\footnote{
\citet{malhotra2002}
have argued on the basis of the distribution of Ly$\alpha$ 
equivalent width in their LAE sample that the LAEs are 
predominantly quite young. 
Thus, the age of LAEs appears to be still under debate. 
However, these two papers may have estimated ages of different 
populations (Of course, differences in the sample selection 
may also be a reason for the inconsistency).
The result of \citet{malhotra2002}
would be for the stellar age of star-forming regions in LAEs, 
since they estimate the age from the rest-frame UV flux 
to which no old star contributes. 
On the other hand, \citet{shapley2001} calculate 
the averaged (luminosity weighted) 
stellar age from the combination of the rest-frame UV
and the rest-frame optical flux, by taking 
advantage of their near-infrared data. Since we 
discuss the averaged feature of LAEs, 
we use here \citeauthor{shapley2001}'s result.
}
A combination of these two findings suggests that
galaxies less luminous in UV continuum are
on the average older.
If we assume here that UV-continuum luminosity correlates
positively with the stellar mass, the suggestion above implies
that less massive galaxies tend to be older,
which is qualitatively consistent with a hierarchical merging
scenario such as CDM models.

  We find a clear clustering signal of the LAEs at $z=4.86\pm0.03$.
The amplitude of angular correlation function (Figure \ref{fig:acorr})
is $A_\omega=29$ arcsec$^{0.8}$, and it is about 40 times
larger than that of
LBGs at $z=3.8\pm0.5$ \citep{ouchi2001a},
$A_\omega=0.71$ arcsec$^{0.8}$.
However, since the selection function in the direction of
redshift
of the LAE sample is about 20 times
narrower than that of LBGs,
the clustering amplitude of $z=4.86$ LAE
in the real space
is found to be comparable to that
of $z\sim4$ LBGs as described below.

\begin{figure*}
\epsscale{0.8}
\plotone{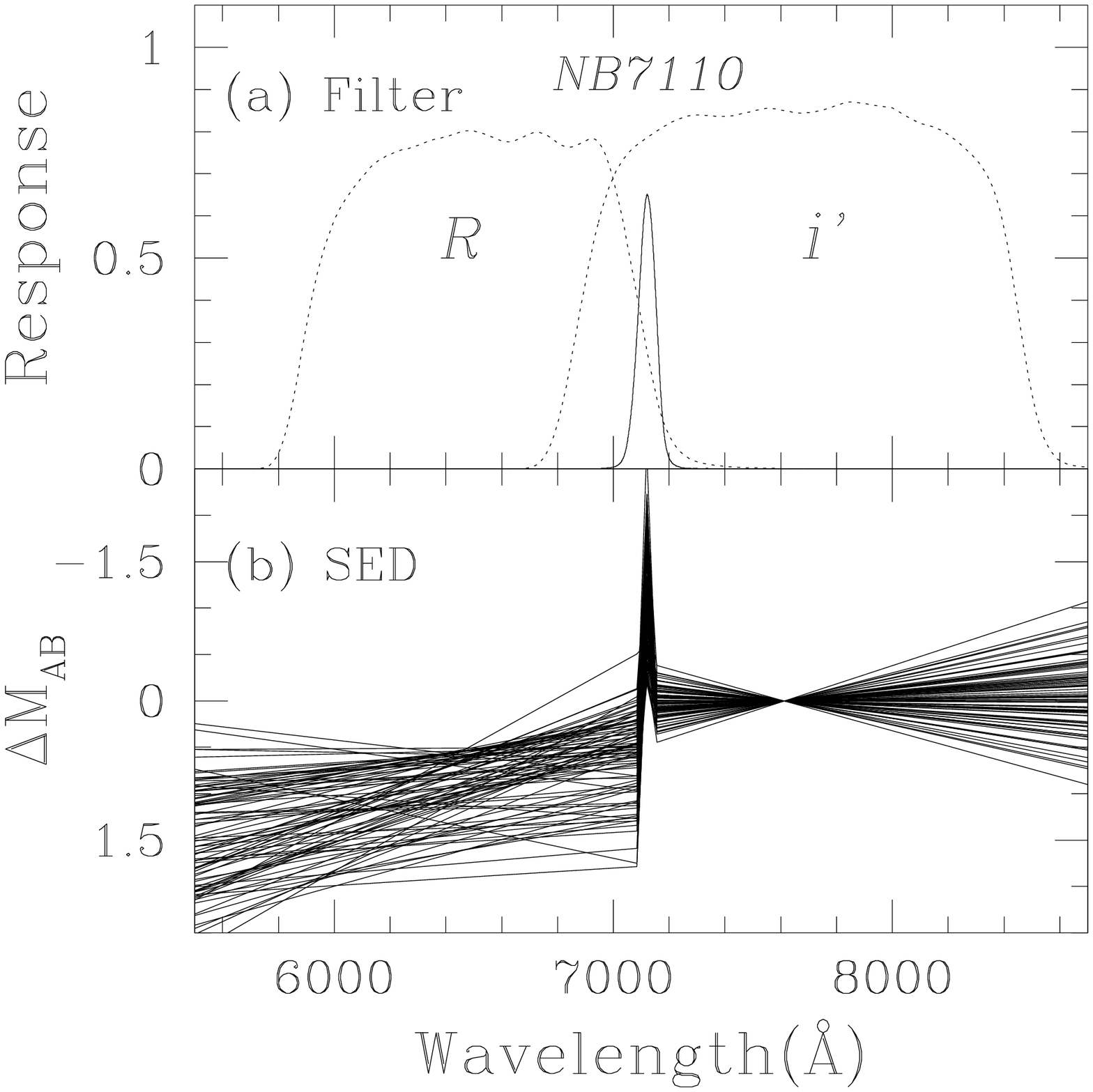}
\caption{
(a)Transmittance of the $NB711$(solid line),
$R$, and $i'$ bands (dotted lines).
(b)SEDs of 87 z=4.86 LAEs. Each solid line corresponds to
the SED of a LAE. $B$,$V$,$R$, $NB711$,
$i'$, and $z'$ magnitudes are simply connected
after normalized so that $i'$ is equal to 0.
     \label{fig:sed_response}}
\end{figure*}

\begin{figure}
\plotone{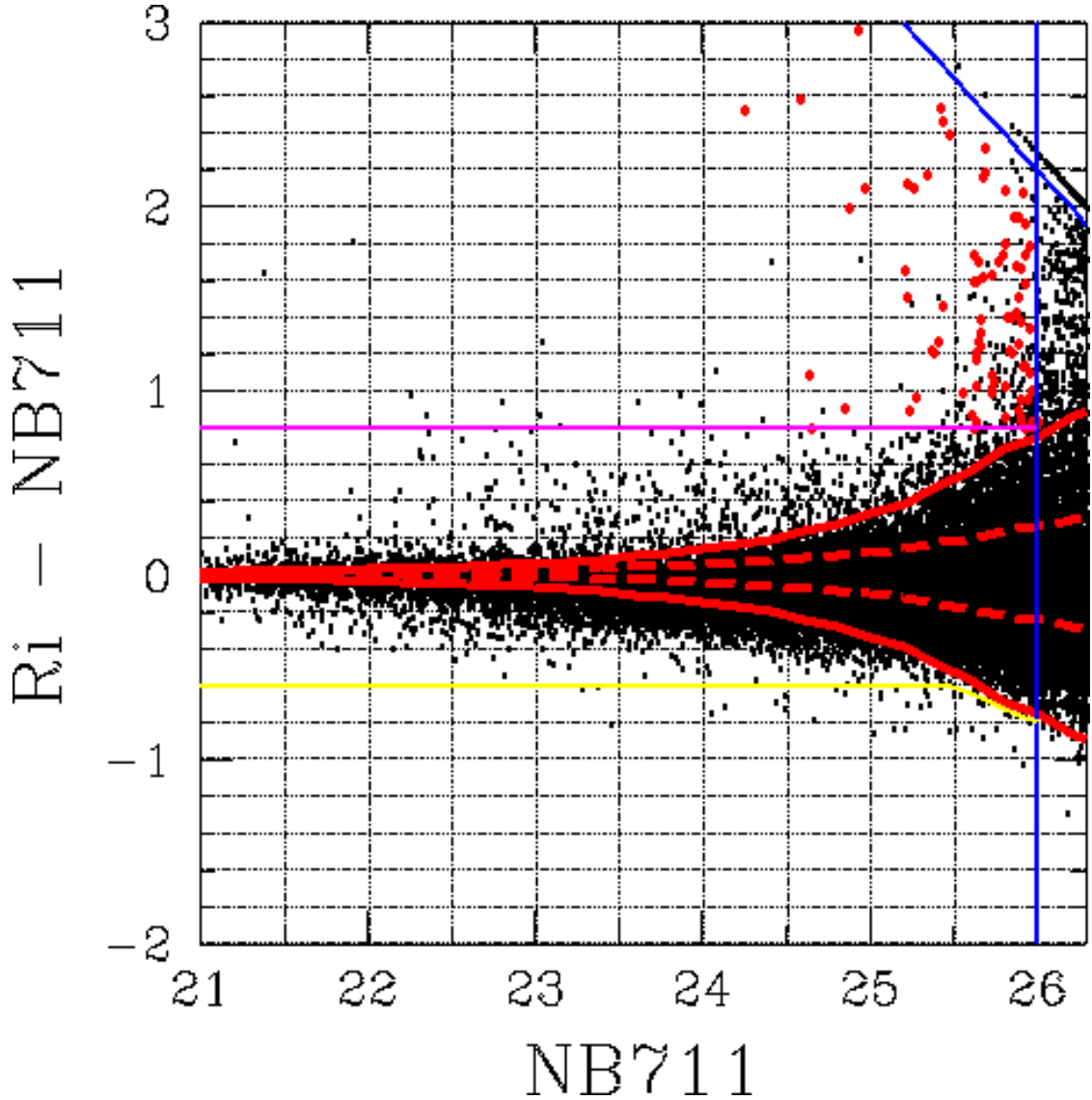}
\caption{
Distribution of all the detected objects in the $Ri'-NB711$
vs. $NB711$ plane, where $Ri\equiv(R+i')/2$.
Red solid (dashed) lines indicate the distribution of
$3\sigma$($1\sigma$) errors in brightness for a
source with a flat ($f_\nu=$const) spectrum.
Blue lines show the detection limits
of $NB711$ and $Ri$,
and the pink line indicates our selection
criterion on $Ri-NB711$ color.
$NB711$-band detected objects
with no continuum emission ($i'>28.1$)
are shown just above the blue line.
Red filled circles are our photometrically
selected $z=4.86$ LAEs which satisfy
all the criteria (see text).
The yellow line indicates the selection criterion
for the absorbers defined in section \ref{sec:selection_sample}.
     \label{fig:cm}}
\end{figure}

\begin{figure}
\plotone{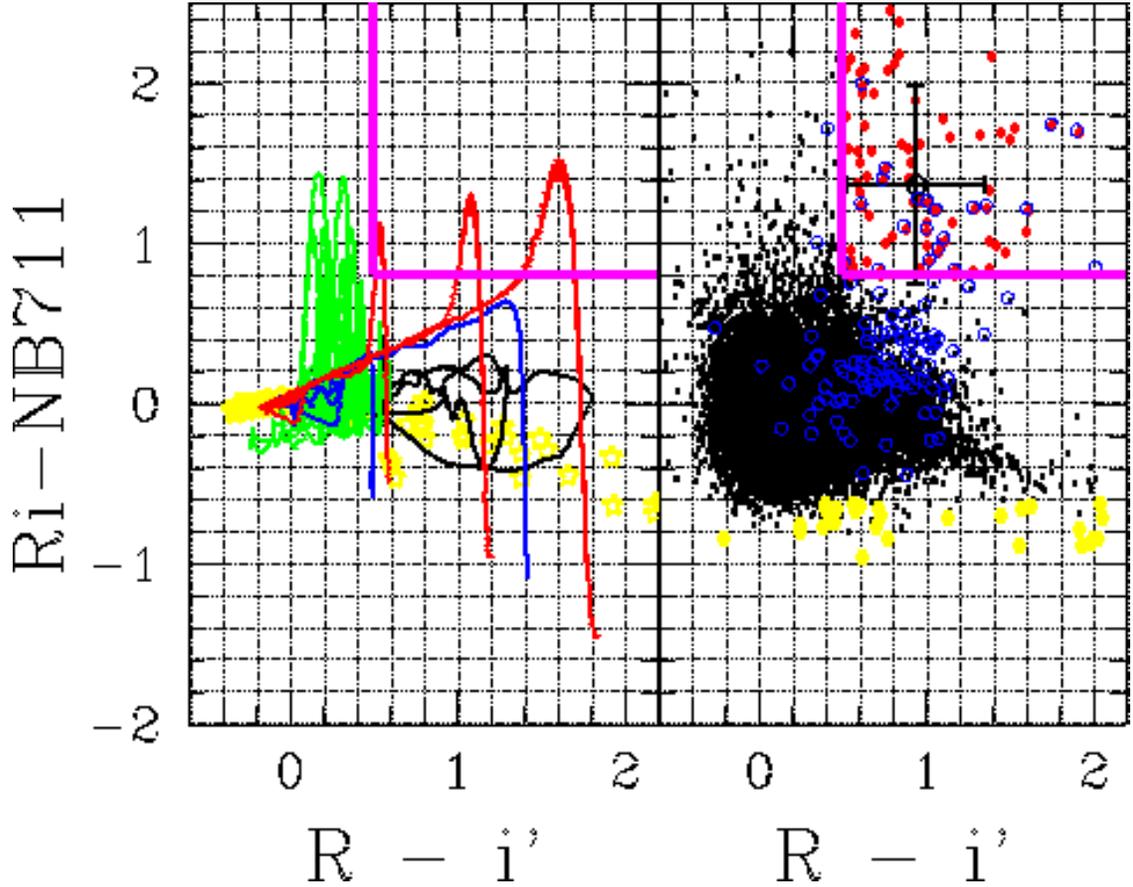}
\caption{Two color diagrams for
continuum color ($R-i'$)
and narrow-band excess color ($Ri-NB711$), where
$Ri$ is a continuum magnitude, $Ri\equiv(R+i)/2$.
({\it left panel})
Tracks of model galaxies at different redshifts.
Red lines indicate model LAE SEDs which are
composite spectra of a 0.03 Gyr single burst
model galaxy (GISSEL96;\citealp{bruzual1993})
and a Lyman $\alpha$ emission
( $EW_{\rm rest}$(Ly$\alpha$)=22 \AA);
from the left to right, three different amplitudes
of IGM absorption are applied:
$0.5\tau_{\rm eff}$, $\tau_{\rm eff}$, and
$1.5\tau_{\rm eff}$, where $\tau_{\rm eff}$ is
the \citeauthor{madau1995}'s \citeyearpar{madau1995}
original median opacity.
The narrow-band excess in each of the peaks in
the red lines
indicates the Lyman $\alpha$ emission
of LAEs at $z=4.86$.
Green lines show 6 templates of
nearby starburst galaxies \citep{kinney1996} up to $z=1.2$,
which are 6 classes of
starburst galaxies with different dust extinction
($E(B-V)=0.0-0.7$). The narrow-band excess
peaks in the green line
of the starbursts correspond to the emission lines of
H$\alpha$ ($z=0.08$), [OIII]($z=0.4$), H$\beta$($z=0.5$),
or [OII]($z=0.9$).
Two blue lines indicate the tracks for model
LBG SEDs without Lyman $\alpha$ emission, for
$0.5\tau_{\rm eff}$, $\tau_{\rm eff}$, respectively.
Black lines show colors of typical
elliptical, spiral, and irregular galaxies \citep{coleman1980}
which are redshifted from $z=0$ to $z=2$. Yellow star marks show
175 Galactic stars given by \citet{gunn1983}, and
24 Kurucz model stars some of which show absorption features
at around 7100\AA.
The pink box surrounding the upper right region is the
selection criteria of our $z=4.86$ LAEs.
({\it right panel}) Colors of the detected objects.
The red circles indicate our LAEs,
while the black dots are the other objects.
All objects satisfying the z$\sim$5 LBG selection
(Ouchi et al. in preparation)
are indicated by blue open circles.
Yellow circles are
37 possible absorption objects 
defined in section \ref{sec:selection_sample}. 
The black open circle
denotes the median color of our 87 LAEs, and the
amplitudes of its error bars indicate the
median photometric errors in the LAEs. Our
LAE selection criteria are also shown by
pink lines.
     \label{fig:cc}}
\end{figure}

\begin{figure}
\plotone{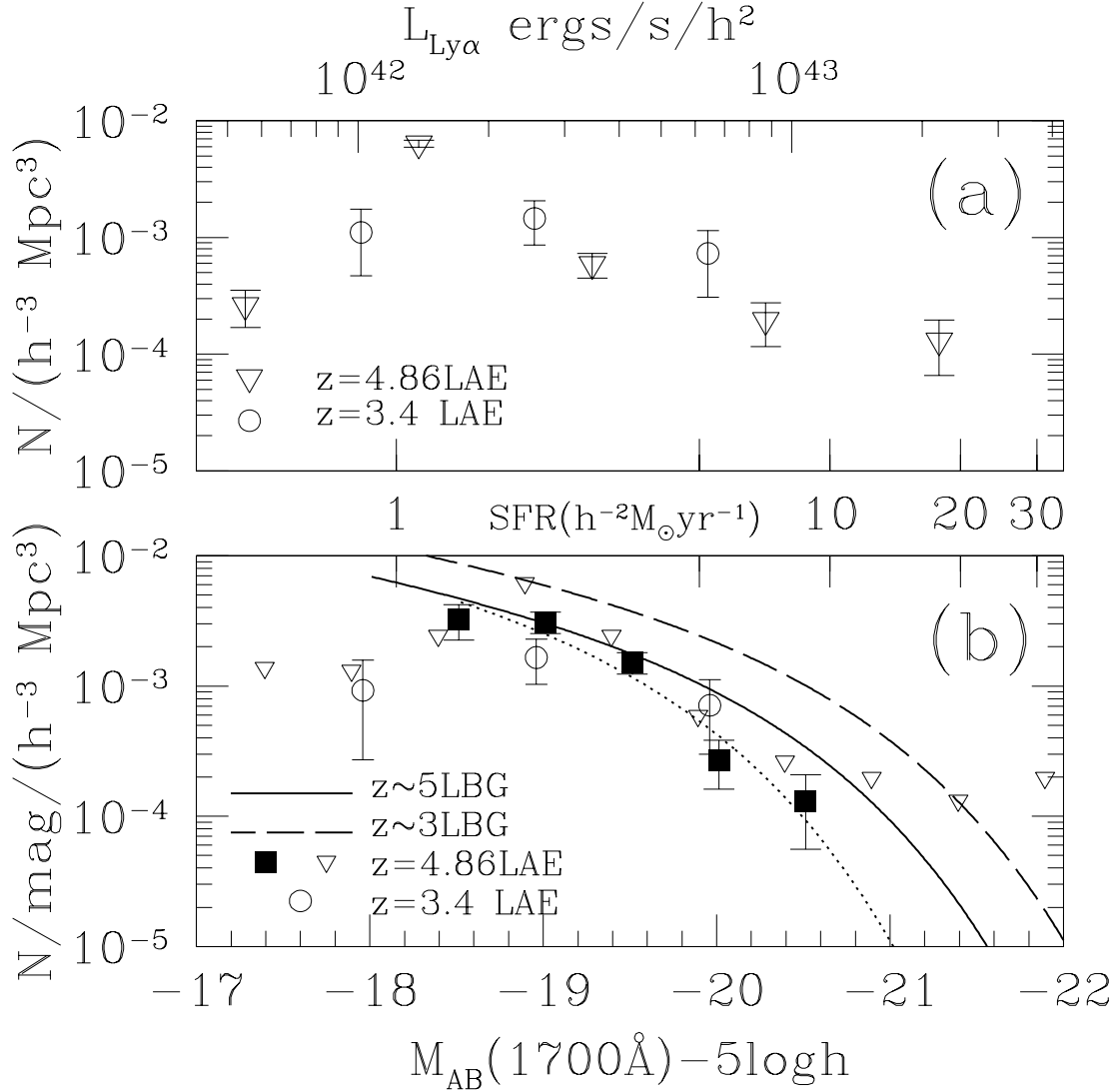}
\caption{
(a)Luminosity functions (LFs) of LAEs
in Ly$\alpha$ luminosity.
Circles are for $z=3.4$ LAEs
with 
$EW_{\rm rest}>17.5$\AA\ selected 
by \cite{cowie1998}.
Inverse triangles are the LF
for $z=4.86$ LAEs 
which are selected from our data
with the same
criteria as for the $z=3.4$ LAEs.
(b)LFs of LAEs and LBGs in UV magnitude.
Filled squares are the best estimated LF for
$z=4.86$ LAEs with 
$EW_{\rm rest}>14$\AA .
The dotted line is the best fit Schechter function
to $z=4.86$ LAEs (with a fixed slope: $\alpha \equiv -1.6$).
The meaning of open circles and open inverse triangles is
the same as
in (a).
The LF for $z=3.4$ LAEs ,shown by inverse triangles,
is probably contaminated by low-$z$ emission-line galaxies,
similar to
\citeauthor{cowie1998}'s \citeyearpar{cowie1998} LF.
Solid and dashed lines are LFs of LBGs at
$z\sim5$ (Ouchi et al. in preparation) and $z\sim3$
(\cite{steidel1999}), respectively.
The $x$ axes of (a) and (b) are marked in star-formation rate ($SFR$).
The $SFR$ in panel (a)
is estimated from Ly$\alpha$ luminosity;
Ly$\alpha$ luminosity is converted into
H$\alpha$ luminosity under the case B approximation
(H$\alpha$=Ly$\alpha /8.7$; \citealp{brocklehurst1971})
and the H$\alpha$ luminosity is translated into
star-formation rate by the relation,
$SFR({\rm M_\odot yr^{-1}}) =
L_{\rm H_{\alpha}}({\rm erg\ s^{-1}})/(1.41\times10^{41})$,
given in \cite{madau1998}.
The star-formation rate in panel (b)
is estimated
from UV-continuum luminosity with the relation,
$SFR({\rm M_\odot yr^{-1}}) = L_{\rm UV}({\rm erg\ s^{-1} Hz^{-1}})
/(8.0\times10^{27})$.
These SFRs are not corrected for dust extinction.
     \label{fig:lumifun} }
\end{figure}

\begin{figure}
\plotone{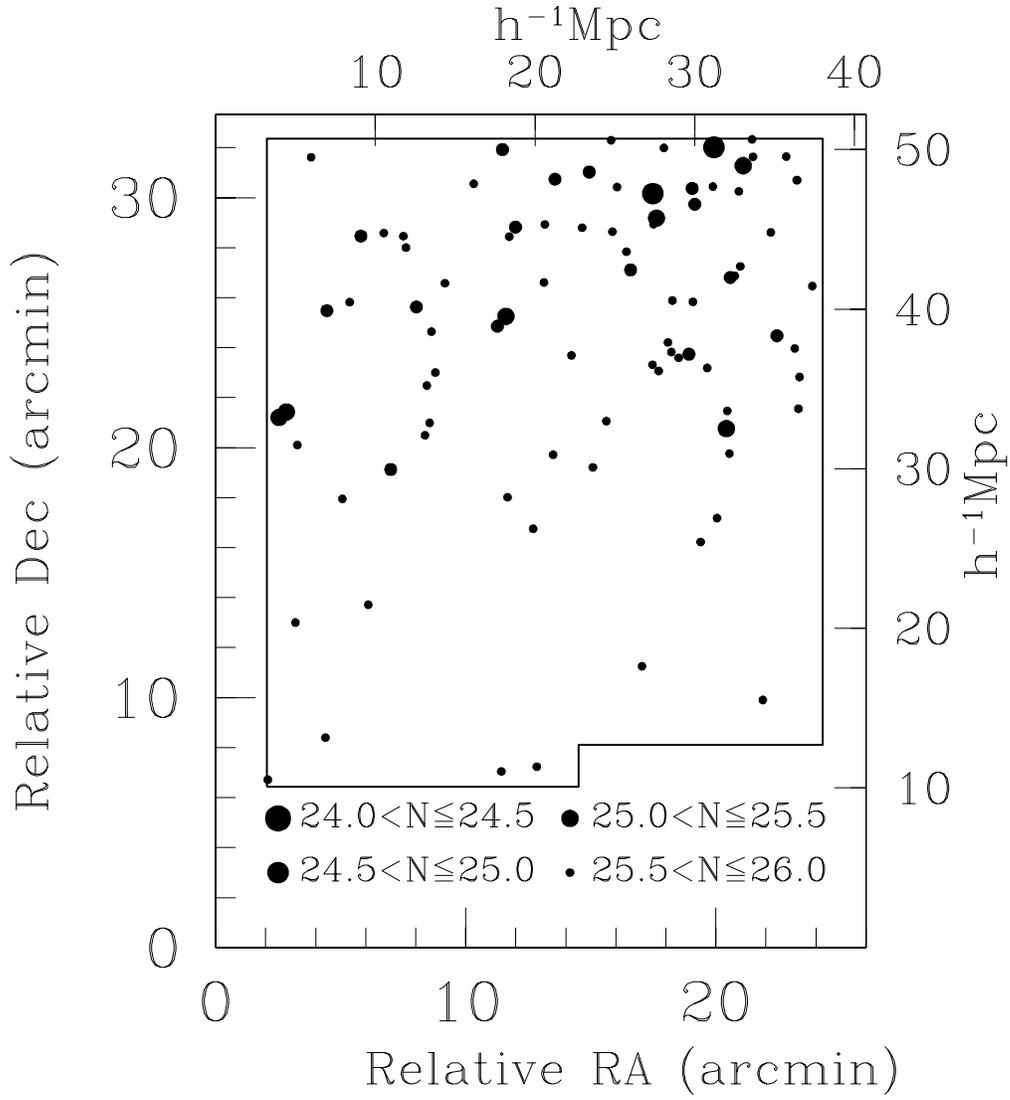}
\caption{ Sky distribution of the photometrically selected $z=4.86$ LAEs.
	Different sizes of circles correspond to different magnitude bins 
	defined in the panel. North is up and east is to the left.
     \label{fig:skydistribution} }
\end{figure}

\begin{figure}
\plotone{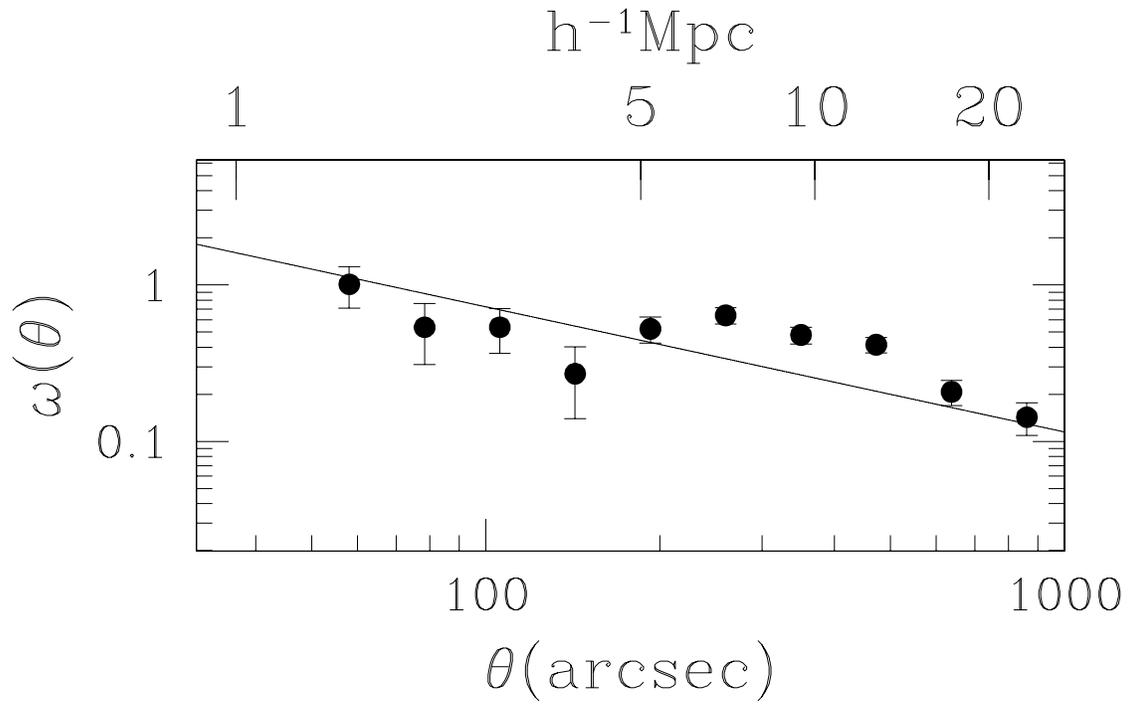}
\caption{Angular correlation function for the 87 LAEs.
     The solid line shows the best fit power law with
     $\omega(\theta)=A_\omega \theta^{-0.8}$
     over the whole range.
     The reduced $\chi^2$ of the fitting is 6.7.
     \label{fig:acorr} }
\end{figure}

The clustering amplitude can be translated into the 
correlation length using Limber equation, given the redshift 
distribution of the sample.
Assuming that the redshift distribution of the LAEs is 
a tophat shape of $z=4.86\pm0.03$, we estimate the correlation 
length, $r_0$, of the LAEs to be 
$r_0=3.5^{+0.3}_{-0.3}h^{-1}$ Mpc
\footnote{
The actual shape of the transmittance of $NB711$ 
can be approximated by a triangle (see Figure \ref{fig:sed_response}a).
We calculate the correlation length assuming
the redshift distribution to be
a triangle whose center and FWHM are
$4.86$ and $0.03$, to obtain a slightly larger value,
$r_0=4.4^{+0.3}_{-0.4}h^{-1}$Mpc.
}.
Foreground contaminations to the sample 
dilute an apparent clustering amplitude of LAEs. 
When the fraction of contaminants is $f$,
the apparent $A_\omega$ value  
can be reduced by a factor of up to $(1 - f)^2$; 
the maximum reduction occurs when the contaminants 
are not at all clustered.
In reality, the contaminants in our sample will be a sum of 
emission-line galaxies at several redshifts, and thus would be 
clustered very weakly on the sky.
If we assume that $f$ is 40\% for our LAE sample 
(see section \ref{sec:selection_sample})
and that contaminants are not clustered, 
we obtain the contamination-corrected
correlation length to be $r_0=6.2^{+0.5}_{-0.5}h^{-1}$Mpc.
This is the maximum $r_0$ value permitted.

These raw and contamination-corrected values of $r_0$ are
larger than that of $z\sim4$ LBGs obtained in \cite{ouchi2001a}; 
$r_0=2.7^{+0.5}_{-0.6}h^{-1}$Mpc for the raw value 
and $r_0=3.3^{+0.6}_{-0.7}h^{-1}$Mpc 
for the contamination-corrected value.
The mean UV luminosity of LAEs in this study is fainter
than that of $z\sim4$ LBGs studied in \citet{ouchi2001a},
since our LAE sample includes galaxies down to $-18.5$ mag,
which is $\sim 1.5$ mag fainter than the faintest galaxy in
\citeauthor{ouchi2001a}'s \citeyearpar{ouchi2001a} LBG sample.
\citet{giavalisco2001} have found in their $z\sim3$ LBGs
that $r_0$ increases with the UV luminosity. If this holds
for LBGs at $z\sim4$ and LAEs at $z=4.86$,
our LAE sample should give a smaller value of $r_0$ than
\citeauthor{ouchi2001a}'s \citeyearpar{ouchi2001a} LBG sample.
However, no such trend is seen; the $r_0$ of $z=4.86$ LAEs is
even larger than that of $z\sim4$ LBGs. This may imply that
$z=4.86$ LAEs have a higher biasing factor against the
underlying dark matter than $z\sim4$ LBGs. (In this discussion,
we assume that clustering properties of galaxies do not significantly
evolve from $z=4.86$ to $z\sim4$).
The general trend that $r_0$ increases 
with halo mass may not necessarily hold for LAEs, 
since their clustering might be 'enhanced' due to, say,
unknown evironmental effects which produce galaxies with 
strong Ly$\alpha$ emission lines. 
However, we cannot rule out at present the possibility that 
the dark halo masses of LAEs are higher than our expectations 
from their luminosities. 

\begin{figure}
\plotone{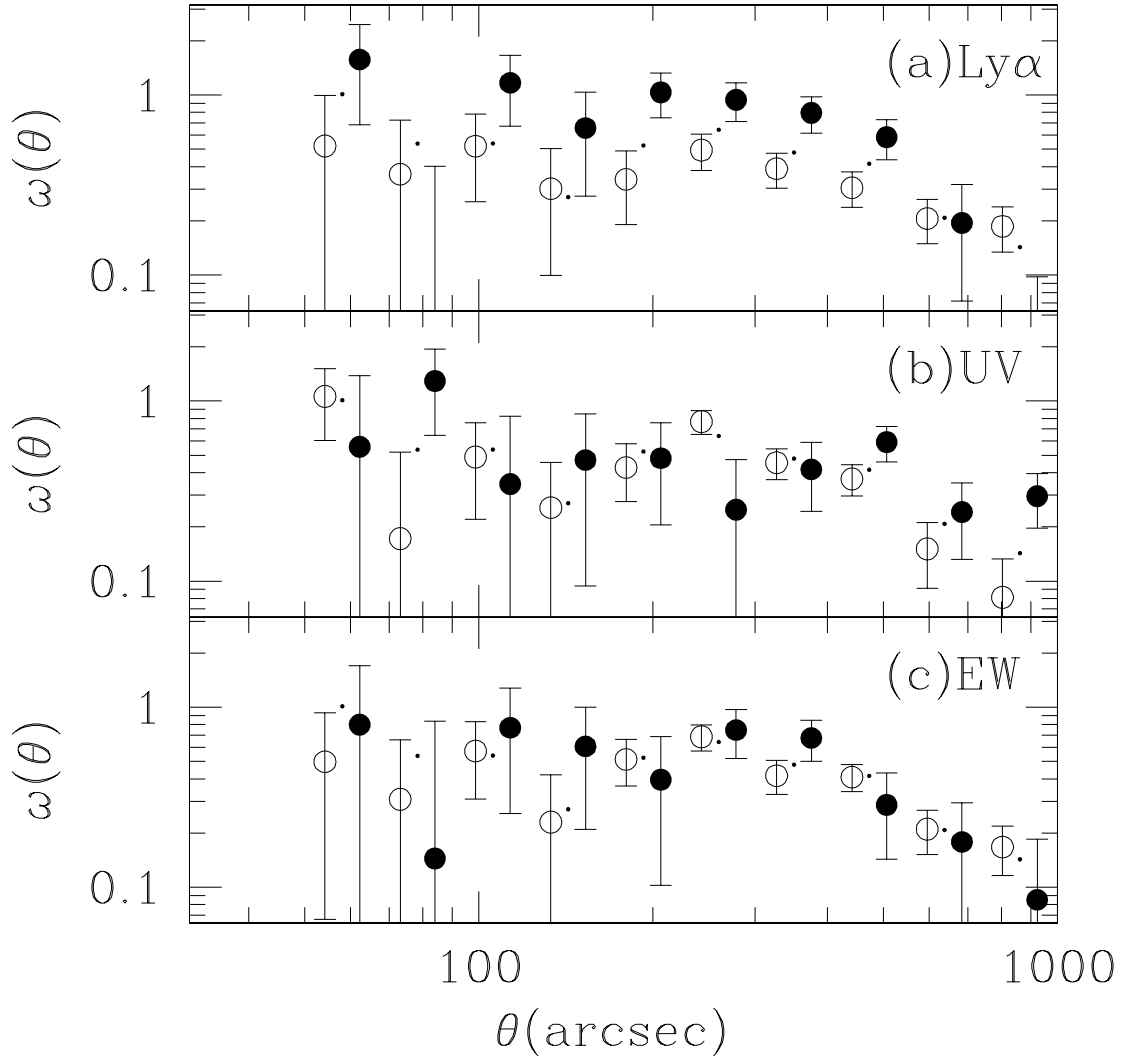}
\caption{ Angular correlation functions for
the sub-samples of 87 $z=4.86$ LAEs.
(a) Ly$\alpha$ luminosity-dependent angular correlation
functions. Filled circles are
for the Ly$\alpha$-bright ($\log L_{\rm Ly\alpha}>42.2$) sample,
while open circles are for the Ly$\alpha$-faint
($\log L_{\rm Ly\alpha}\leq42.2$ ) sample.
Dots indicate
the angular correlation function of the whole 87 LAEs
shown in Figure \ref{fig:acorr}.
The filled (open) circles
are shifted by $0.03$dex ($-0.03$dex)
along the abscissa for clarity.
(b) Same as (a), but for UV-continuum luminosity.
Filled circles are for the UV-continuum bright ( $M_{1700} <-19$)
sample, and open circles are for the UV-continuum faint ($M_{1700}\leq -19$ )
sample.
(c) Same as (a), but for observed equivalent width ($EW_{\rm obs}$) of
Ly$\alpha$ emission.
Filled circles are for the $EW_{\rm obs}$ large
( $EW_{\rm obs}> 270$\AA ) sample,
and open circles are for the $EW_{\rm obs}$ small
( $EW_{\rm obs}\leq 270$\AA ) sample.
     \label{fig:acorr_seg} }
\end{figure}

The LAEs in our sample have nearly the same dynamic range
both in Ly$\alpha$ luminosity
($41.85 \le \log L_{\rm Ly \alpha} \le 42.65$ ) and
in UV-continuum luminosity ($M_{1700}=-18.5$ to $-20.5$).
Figure \ref{fig:acorr_seg} shows, however, that the 
difference of clustering amplitude is seen 
not for UV sub-samples but only for 
Ly$\alpha$ sub-samples. Hence the degree of clustering
is probably more strongly dependent on Ly$\alpha$ luminosity
than UV-continuum luminosity.
This result is consistent with the fact found above that
$z=4.86$ LAEs show stronger clustering amplitude than 
$z\sim5$ LBGs in spite of the former being fainter 
in UV-continuum luminosity.

\citet{shapley2001} have found in their $z\sim3$ LBG sample that
older LBGs (in terms of mean stellar age) have a stronger Ly$\alpha$
emission. If this is also the case for LAEs,
LAEs with a stronger Ly$\alpha$ emission will be older on the average.
This is consistent with our finding that LAEs
more luminous in Ly$\alpha$ emission are more strongly clustered,
if we assume that the biasing factor is higher for older galaxies.

We have found that LAEs dominate at the fainter part of
the UV LF of galaxies at $z\sim5$. We have also found a clear
clustering signal in our $z=4.86$ LAE sample.
\citet{pascarelle1996} have made the hypothesis,
from their findings of $z=2.4$ LAEs being compact,
that LAEs may be sub-galactic clumps
located in groups or sheets
of collapsed dark matter.
Our two findings seem to be consistent
with the hypothesis.

\acknowledgments
We would like to thank the Subaru Telescope staff
for their invaluable help in commissioning the Suprime-Cam
that made these difficult observations possible.
We thank the referee, James E. Rhoads, for his detailed comments
that improved this article.
M. Ouchi, H. Furusawa, F. Nakata, and Y. Shioya acknowledge
support from the Japan Society for the
Promotion of Science (JSPS) through JSPS Research Fellowships
for Young Scientists.




\clearpage


\clearpage

\clearpage

\clearpage

\clearpage

\clearpage

\clearpage

\end{document}